\documentclass[aps,prl,twocolumn,superscriptaddress]{revtex4-1}
\usepackage{graphicx}
\usepackage{mathrsfs}
\usepackage{bm}
\usepackage{amsmath}
\usepackage{color}
\usepackage{dcolumn}
\usepackage{dsfont}
\usepackage{amssymb}
\usepackage{tabularx}
\usepackage{array}
\usepackage{float}
\usepackage{colordvi}
\usepackage{verbatim}
\usepackage{hyperref}

\begin{document}
\title{Realization of the Non-endpoint Majorana Bound States in an Extended Kitaev Chain}

\author{Xuan Zhang}
\affiliation{College of Physics, Hebei Normal University, Shijiazhuang 050024, China}
\author{Cheng-Ming Miao}
\affiliation{College of Physics, Hebei Normal University, Shijiazhuang 050024, China}
\author{Qing-Feng Sun}
\email[]{sunqf@pku.edu.cn}
\affiliation{International Center for Quantum Materials, School of Physics, Peking University, Beijing 100871, China}
\affiliation{Hefei National Laboratory, Hefei 230088, China}
\author{Ying-Tao Zhang}
\email{zhangyt@mail.hebtu.edu.cn}
\affiliation{College of Physics, Hebei Normal University, Shijiazhuang 050024, China}

\begin{abstract}
We study the energy levels and transport properties of an extended Kitaev chain with a phase gradient.
It is demonstrated that the hopping phase difference can effectively induce the generation of Majorana bound states, which are located at the non-endpoint sites of the chain.
The number and positions of the non-endpoint Majorana bound states can be modulated by the hopping phase difference and initial hopping phase, respectively.
In addition, we propose a protocol to realize topological braiding operation by exchanging the positions of two Majorana bound states in the extended Kitaev ring. Furthermore, we also implement the braiding of any two of the multiple Majorana bound states in the extended Kitaev double rings.
\end{abstract}

\maketitle

\emph{Introduction}.---
Majorana bound states (MBSs) are zero-energy quasiparticles
obeying non-Abelian statistics~\cite{Kitaev2001, Nayak2008, Wilczek2009, Alicea2012, Leijnse2012, Beenakker2013}, which means that exchanging
two MBSs results in a transformation depending on the order of
the exchange. These properties make them as ideal candidates for
the realization of topological quantum computation~\cite{Nayak2008, Alicea2011}.
There have been many theoretical proposals suggesting that
MBSs exist in the vortex core of p-wave topological superconductors
or at the ends of one-dimensional topological superconductors~\cite{Kitaev2001, Ivanov2001, Lutchyn2010, Oreg2010, Alicea2012, Nadj2013, Hell2017}.
Despite experiments show signatures expected for MBSs,
the various possibilities makes the interpretation challenging~\cite{Bagrets2012, Rainis2013, Liu2012, Liu2017, Moore2018}. Therefore, detecting and braiding MBSs still remains important
and open challenge. In a rigorous one-dimensional system,
two MBSs cannot bypass each other for braiding.
Thus, some artificial two-dimensional platforms have been proposed
to avoid mixing MBSs, such as introducing T/Y-junctions, quantum dots,
and so on~\cite{Karzig2016, Yan2019, Zhou2019, Mishmash2020, Manousakis2020}.
All these schemes inevitably require the aid of other chains
or dots, as well as are difficult to extend to braiding multiple MBSs,
since the positions of the MBSs are
in two endpoints of the one-dimensional chains.

\begin{figure}
  \centering
  \includegraphics[width=8.4cm,angle=0]{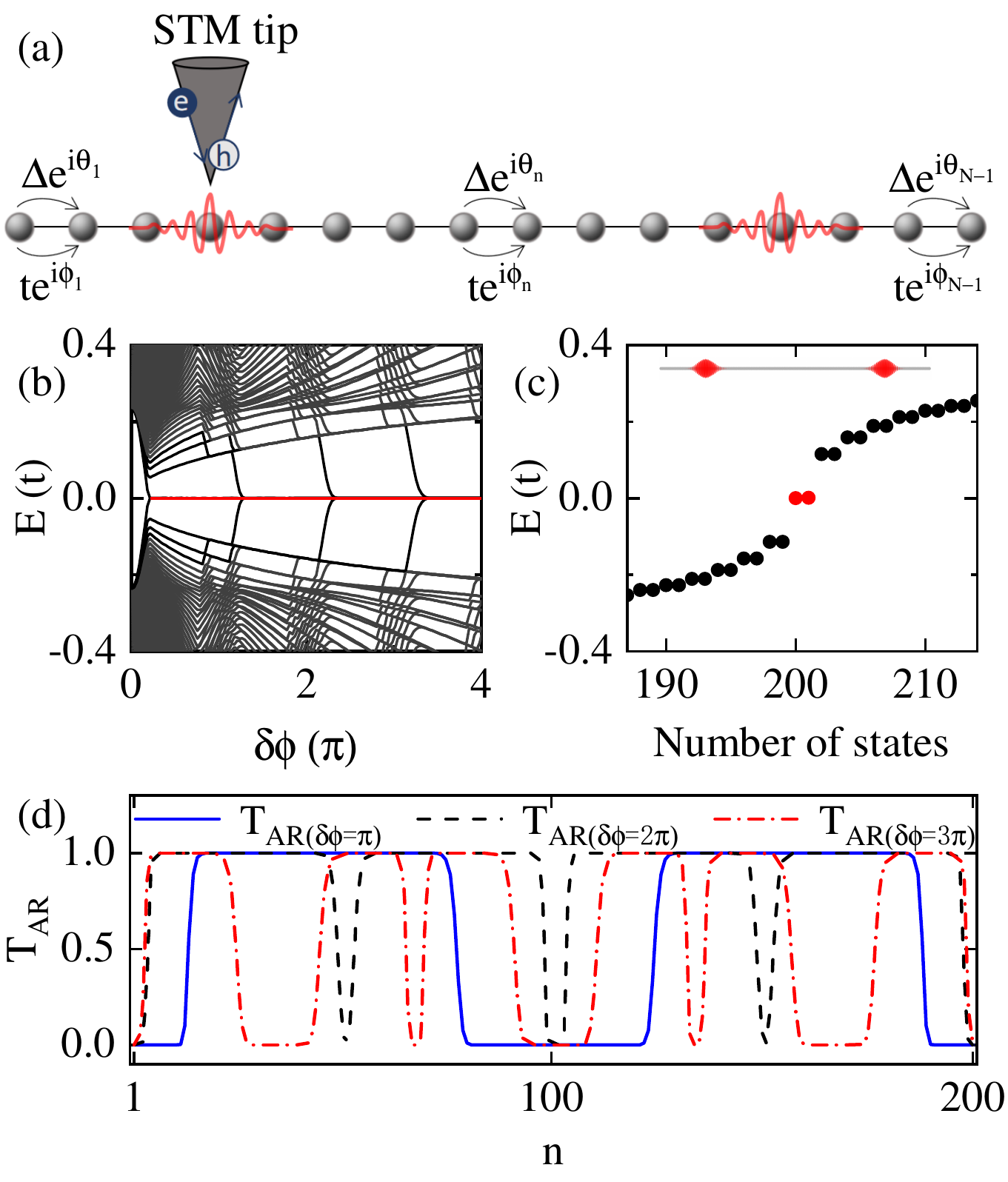}
  \caption{(a) Schematic of an extended Kitaev chain with site-dependent hopping and superconducting phase.
  The STM tip can scan along whole the chain.
  The red curves represent MBSs.
  (b) Band structure for the Kitaev chain as a function of hopping phase difference $\delta\phi$. The zero-energy in-gap states are denoted by the red lines.
  (c) Energy levels of Kitaev chain with $\delta\phi=\pi$.
  The red dots represent the zero-energy MBSs
  whose probability distributions are plotted in the inset.
  (d) Andreev reflection coefficient $T_{AR}$ as a function of STM tip connected
  sites $n$ with $\delta\phi=\pi$ (blue solid line), $2\pi$ (black dashed line) and $3\pi$ (red dotted-dashed line).
  The energy of the incident electron is $E =0$, and the other parameters are set to be $\mu=1$, $t=1$, $\Delta=1$, $\phi_1=0.5\pi$, $\theta_1=0$, $\delta\theta=0$, and $N=200$.}
  \label{fig1}
\end{figure}

In this letter, we investigate the energy levels and transport properties
of an extended Kitaev chain with a phase gradient.
It is found that MBSs are bound at non-endpoint sites of
the extended Kitaev chain in the presence of a hopping phase difference,
and their number and positions can be adjusted by the hopping phase difference and initial hopping phase. Furthermore, we exchange the positions of MBSs to complete the quantum braiding process in an extended Kitaev ring, which is always topologically protected by an energy gap.

\emph{Model Hamiltonian}.---We consider an extended Kitaev chain, in which the hopping phases and $p$-wave superconducting phases are site-dependent and linearly increasing, as shown in Fig.~\ref{fig1}(a).
The Hamiltonian of the extended Kitaev chain can be described as \cite{Kitaev2001}:
\begin{align}
 H_{C}=&-\mu\sum_{n=1}^{N}c_{n}^{\dag}c_{n}-t\sum_{n=1}^{N-1}e^{i\phi_n}(c_{n}^{\dag}c_{n+1}+h.c.)\nonumber \\
&+\Delta \sum_{n=1}^{N-1}e^{i \theta_n}(c_{n}c_{n+1}+h.c.),
\label{eq1}
\end{align}
where $c_{n}^{\dag}$ ($c_{n}$) is fermionic creation (annihilation) operator at site $n$, and the total number of the site is chosen as $N$.
$\mu$ is the chemical potential, $t$ is the nearest-neighbor hopping strength amplitude, and $\Delta$ is the $p$-wave pairing amplitude.
The phases of hopping and superconducting between site $n$ and site $n+1$ are $\phi_{n}=\phi_{1}+\frac{n-1}{N -1}\delta\phi$ and $\theta_{n}=\theta_{1}+\frac{n-1}{N -1}\delta\theta$, respectively. $\delta\phi=\phi_{N-1}-\phi_{1}$ ($\delta\theta=\theta_{N-1}-\theta_{1}$) is the phase difference between the first and last hopping (superconducting) phases, and $\phi_{1}$ ($\theta_{1}$) is the initial hopping (superconducting) phase.

\emph{Realization of the non-endpoint MBSs}.---In Fig.~\ref{fig1}(b), we calculate the energy levels of the extended Kitaev chain as a function of hopping phase difference $\delta\phi$.
One can see that the twofold degenerate zero-energy states emerge at $\delta\phi=0.25\pi$.
The degeneracy of the zero-energy states increases in pairs
with the increase of the hopping phase difference.
Furthermore, we plot the energy levels with $\delta\phi=\pi$ in Fig.~\ref{fig1}(c).
It is shown that there are two zero-energy in-gap states (red dots),
the probability distribution of which is highlighted in the inset of Fig.~\ref{fig1}(c).
One can see that the zero-energy in-gap states are not located
at the endpoint positions of the superconducting chain.
It is worth noting that the zero-energy in-gap states are the MBSs
because they are composed of electrons and holes each contributing half probability.
While for the hopping phase difference $\delta\phi=2\pi$ and $\delta\phi=3\pi$, the four and six degeneracy MBSs appear, the wave function of which is distributed at non-endpoint position among the extended Kitaev chain,
see Fig. S1 in Supplemental Material~\cite{supp}.
Our results demonstrate that non-endpoint MBSs can be induced in the extended Kitaev chain with the hopping phase difference. In addition, the number of non-endpoint MBSs can be adjusted by the magnitude of the hopping phase difference.

It is well known that a MBS can lead to a perfect Andreev reflection ($T_{AR}=1$), a process of converting an incoming electron into an outgoing hole~\cite{Zhang2017, Li2020, Zhuang2022, Miao2022}.
Thus the scanning tunneling microscope (STM) spectroscopic mapping
can be taken as the single Majorana detector.
To detect the location of the MBSs, we employ an STM tip to scan each lattice site $n$ of the extended Kitaev chain [see Fig.~\ref{fig1}(a)].
We employ the nonequilibrium Green's function method~\cite{addr1,Sun2009} to calculate  the Andreev reflection coefficient of the extended Kitaev chain connected by the STM tip.
The Andreev reflection coefficient can be described as
$T_{\textrm{AR}}(E)=Tr[\Gamma_{ee}G_{eh}^r\Gamma_{hh}G_{he}^a],$
where $e$ and $h$ represent electron and hole, respectively~\cite{addr1,Sun2009}.
The linewidth functions is $\Gamma (E)=i[\Sigma^{r}-\Sigma^{a}]$. $\Sigma^{r}$ is the self-energy due to the coupling between the STM tip and the Kitaev chain. $G^{r}(E)=[G^{a}(E)]^{\dagger}=[E -H_{\textrm{BdG}}-\Sigma^r]^{-1}$ are the retarded and advanced Green's functions.

In Fig.~\ref{fig1}(d), we plot the Andreev reflection coefficient $T_{AR}$ as a function of connection site $n$ of the extended Kitaev chain with different hopping phase differences $\delta\phi=\pi$, $2\pi$ and $3\pi$.
Figure~\ref{fig1}(d) shows that the perfect Andreev reflection coefficient $T_{AR}=1$ occurs when the STM tip scans the non-endpoint sites of the chain, while no Andreev reflection with $T_{AR}=0$ occurs when scanning with STM tip to the end sites of the chain.
The number and position of the perfect reflection coefficient platform $T_{AR}=1$ are consistent with the number and distribution position of zero-energy non-endpoint bound states in the insets of Fig.~\ref{fig1}(c) and Fig. S1 in Ref.~\cite{supp}.
The above results provide lateral evidence that the zero-energy non-endpoint bound states are MBSs whose numbers can be modulated by the hopping phase difference.

\begin{figure}
	\centering
	\includegraphics[width=8.4cm,angle=0]{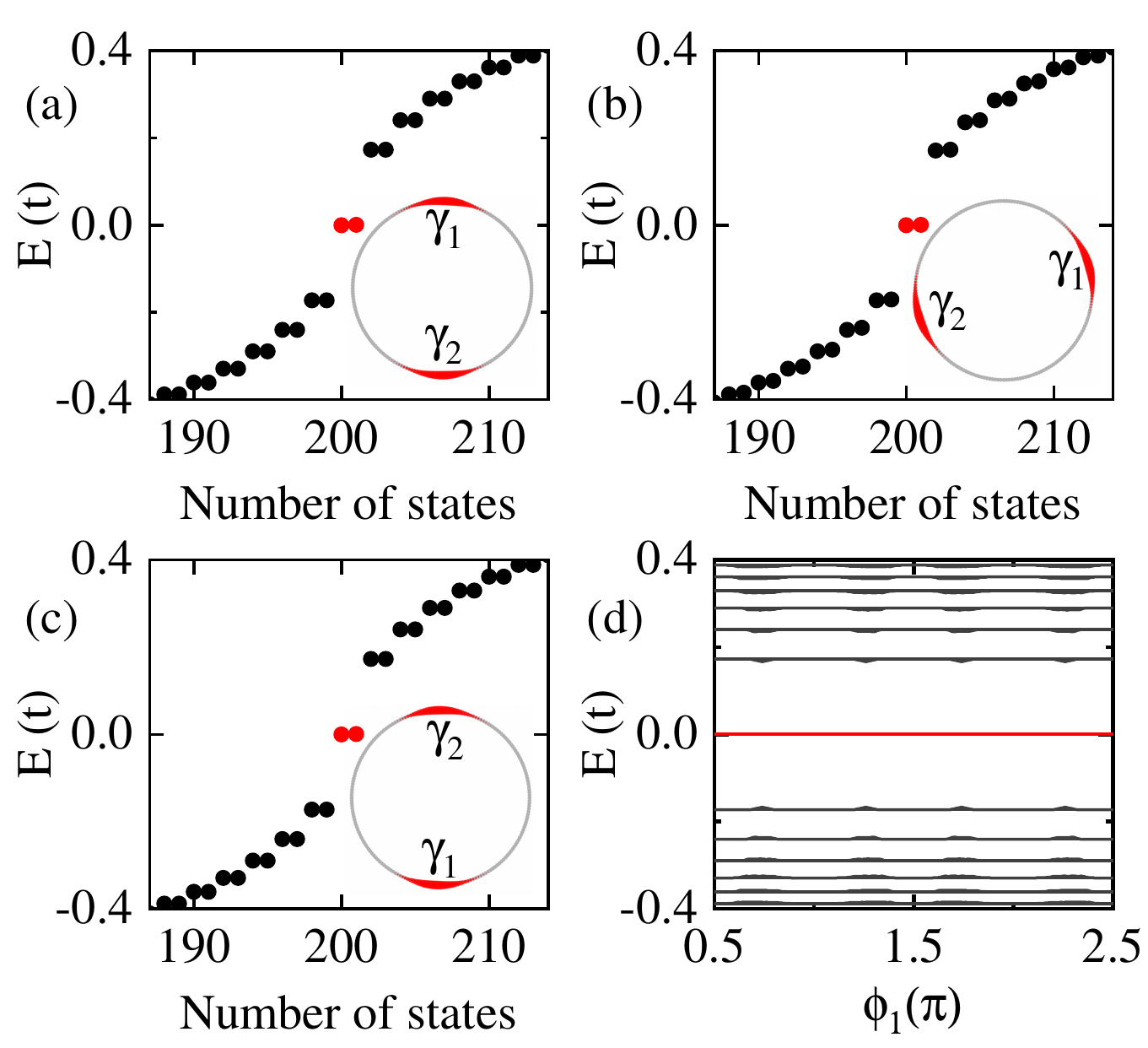}
	\caption{(a)-(c) Energy levels of Kitaev ring with different initial hopping phases $\phi_1=0.5\pi$ for (a), $\phi_1=0.7\pi$ for (b) and $\phi_1=\pi$ for (c). The red dots represent the in-gap MBSs, the probability distribution of the MBSs is plotted in the inset.
	(d) Band structure for the Kitaev ring as a function of the initial hopping phase $\phi_1$. Red lines indicate the Majorana zero-energy bands,
	which always remain separate from the other bands (black lines).
	The other parameters are set to be $\mu=1.4$, $t=1$, $\Delta=1$, $\delta\phi=\pi$, $\theta_1=0$, $\delta\theta=\pi$ and $N=200$.}
	\label{fig2}
\end{figure}

\emph{Braiding of the non-endpoint MBSs}.---Naturally,
we connect the first and last sites of the one-dimensional chain with non-endpoint MBSs into a superconducting ring,
as shown in Fig. S2 in Ref.~\cite{supp}.
The Hamiltonian for the extended Kitaev ring can be written by adding two terms to Eq.~({\ref{eq1}}):
\begin{align}
H_{R}=H_{C}-t e^{i\phi_N}c_{N}^{\dag}c_{1}+\Delta e^{i\theta_N}c_{N}c_{1}+h.c.,
\label{eq2}
\end{align}
where the adding two terms represent the hopping and the pairing interaction between the first and last sites of the chain.
The hopping (superconducting) phase between site $n$ and site $n+1$ is
rewritten as $\phi_{n}=\phi_{1}+\frac{n-1}{N}\delta\phi$ ($\theta_{n}=\theta_{1}+\frac{n-1}{N}\delta\theta$), in which the phase difference between the first and last hopping (superconducting) phases is denoted by $\delta\phi=\phi_{N}-\phi_{1}$ ($\delta\theta=\theta_{N}-\theta_{1}$). The other parameters are consistent with those in Eq.~({\ref{eq1}}).

In Figs.~\ref{fig2}(a)-~\ref{fig2}(c), we calculate the energy levels of the Kitaev ring with different initial hopping phases $\phi_1$.
Figure~\ref{fig2}(a) shows that two zero-energy in-gap MBSs (red dots)
are still present in the extended Kitaev ring for $\phi_1=0.5$.
Since MBSs are at the non-endpoints of the extended Kitaev chain,
they do not coupled and disappear when the chain is linked head-to-end.
The two zero-energy eigenvalues of the system are denoted as $\varepsilon_{1,2}$, with the corresponding eigenstates $|\phi_{\varepsilon_{1,2}}\rangle$. Theoretically, the isolated Majorana fermions $\gamma_1$ and $\gamma_2$ can be obtained by combinations of these eigenstates, i.e., $|\phi_{\gamma_{1,2}}\rangle = (1/\sqrt{2})(|\phi_{\varepsilon_1}\rangle\pm|\phi_{\varepsilon_2}\rangle)$.
At $\phi_1=0.5$, $\gamma_1$ and $\gamma_2$ arise at the upper and lower positions of the Kitaev ring, respectively [see the inset of Fig.~\ref{fig2}(a)].
These two MBSs can be considered as our initial state
in the topological braiding processing.
By increasing the initial phase to $\phi_1=0.7$,
the two zero-energy in-gap MBSs remain [see Fig.~\ref{fig2}(b)],
but the positions of $\gamma_1$ ($\gamma_2$) move to the upper right (lower left) sites of the Kitaev ring, as shown in the inset of Fig.~\ref{fig2}(b).
It indicates that the positions of the MBSs can be adjusted by changing the initial hopping phase.
Until $\phi_1=\pi$, two MBSs $\gamma_1$ and $\gamma_2$ arise at the lower and upper positions of the Kitaev ring, respectively.
The spatial positions of $\gamma_1$ and $\gamma_2$ are mutually swapped as displayed in the inset of Fig.~\ref{fig2}(c).
The braiding operation can be represented by the unitary operator $U(\gamma_1,\gamma_2)=exp(-\frac{\pi}{4}\gamma_1\gamma_2)$. The MBSs can be transformed as $\gamma_1 \rightarrow \gamma_2$ and $\gamma_2 \rightarrow -\gamma_1$ \cite{Ivanov2001}.
To implement an entire topological braiding process,
we continue to increase the initial hopping phase until $\phi_1=1.5\pi$
and show the entire braiding process in Fig. S3 in the Supplemental Material \cite{supp}. In the Video 1 in Ref.~\cite{supp}, we show
the animation of the entire braiding process with the increase of $\phi_1$ from $0.5\pi$ to $1.5\pi$.
In this entire braiding process, the spatial positions of $\gamma_1$
and $\gamma_2$ can be
swapped twice and the system returns to its initial state.
After the exchange process, both MBSs $\gamma_1$ and $\gamma_2$
accumulate a $\pi$ Berry phase and experience a sign flip,
with $\gamma_1 \rightarrow -\gamma_1$ and $\gamma_2 \rightarrow -\gamma_2$. It is worth noting that the positions of $\gamma_1$ and $\gamma_2$ are exchanged without spatial collision.
In order to observe whether the two MBSs are excited throughout the entire braiding process, we plot the energy levels of the extended Kitaev ring as a function of the initial hopping phase $\phi_{1}$ [see Fig.~\ref{fig2}(d)].
One can see that the zero-energy MBSs (red lines)
remain stable throughout the variation of $\phi_{1}$.
The isolated zero-energy MBSs can completely prevent mixing with other states (black lines) by an energy gap. Thus, our results validate the stability of the braiding process in the extended Kitaev ring.
Moreover, the two MBSs can also be exchanged counterclockwise by setting $\phi_1$ from $1.5\pi$ to $0.5\pi$.

By using the above extended Kitaev ring, we can also realize
the braiding process of multiple MBSs.
For example, in order to obtain an efficient braiding process of four MBSs,
we set the phase differences $\delta\phi=\delta\theta=2\pi$.
By increasing the initial hopping phase $\phi_1$,
the four MBSs rotate in the ring.
For $\phi_1 \in [0, 2\pi]$, the spatial positions of $\gamma_1$ ($\gamma_4$) and $\gamma_3$ ($\gamma_2$) are swapped twice and the system returns to its initial state, see Figs. S4(a-e) in the Supplemental Materials \cite{supp}. The combination of the electron tunneling and rotation can identify a novel braiding operator \cite{Park2020}.
Furthermore, it is validated that the stabilized braiding process of four MBSs was obtained by tuning the initial hopping phase [see Fig. S4(f) in Ref.~\cite{supp}].

\begin{figure}
	\centering
	\includegraphics[width=8.4cm,angle=0]{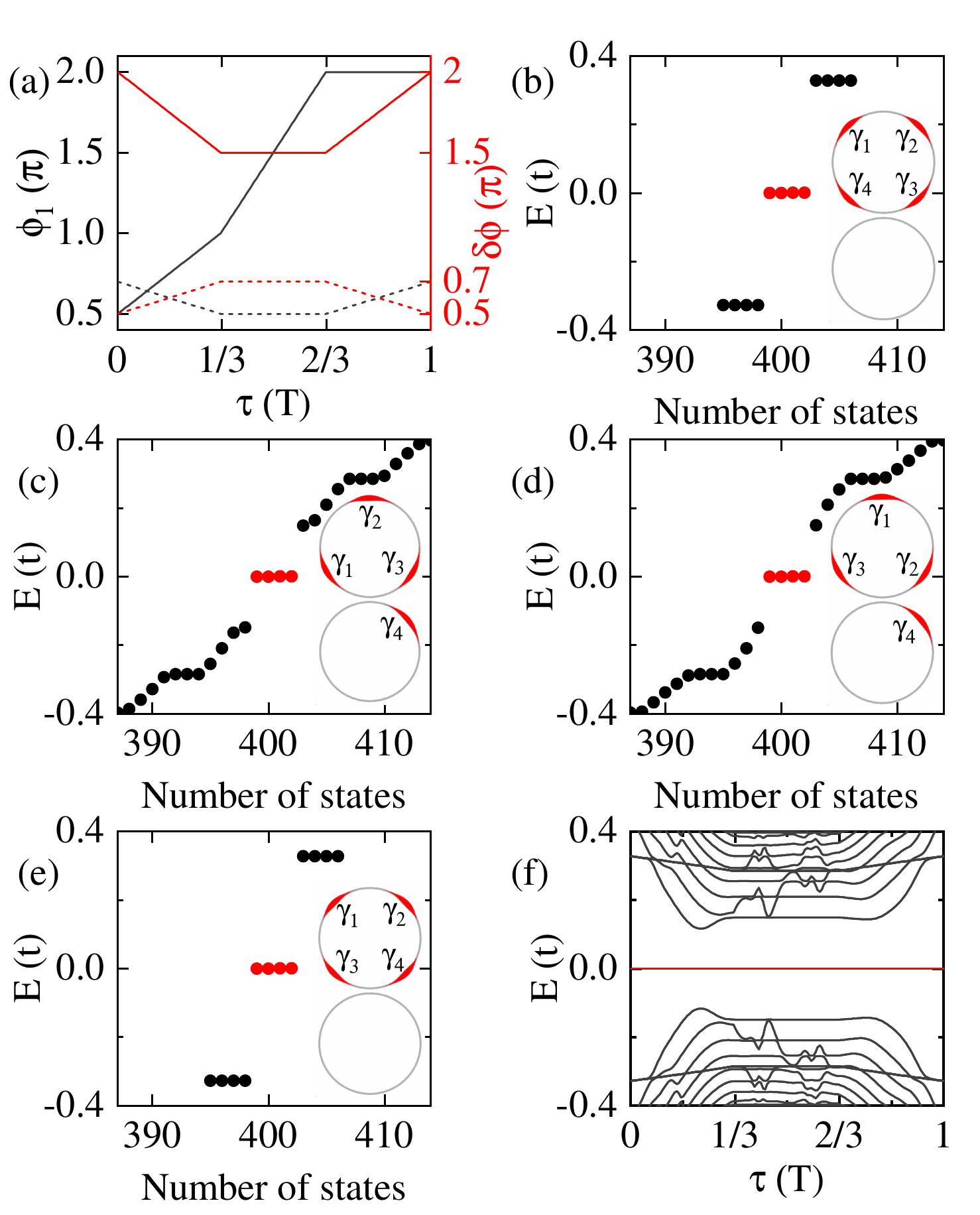}
	\caption{(a) Initial hopping phase $\phi_{1}^{1,2}$ (black lines) and hopping phase difference $\delta\phi^{1,2}$ (red lines) vs time $\tau$ in the Kitaev double rings. The solid and dashed lines show the variation of the parameters in rings $R_1$ and $R_2$, respectively.
(b)-(e) Energy levels of the Kitaev double rings with different time $\tau=0$ for (b), $\tau=T/3$ for (c), $\tau=2T/3$ for (d) and $\tau=T$ for (e). The red dots represent the in-gap MBSs, the probability distribution of the MBSs is plotted in the inset. (f) Band structure for the Kitaev double rings as a function of time $\tau$. Red lines indicate the zero-energy MBSs, which always remain separate from the other states (black lines). The parameters of the rings $R_1$ and $R_2$ are set to be $\mu_{1}=1.4$, $\theta_1^{1}=0$, $\delta\theta^{1}=1.5\pi$, $N_{1}=200$, $\mu_{2}=0.6$, $\theta_1^{2}=1.5\pi$, $\delta\theta^{2}=0.5\pi$, and $N_{2}=200$. The other parameters of the double rings are the same as $t=1$, $t_1=1$ and $\Delta=1$.}
	\label{fig3}
\end{figure}

To enable more elementary gate operations,
we design the Kitaev double rings composed of Kitaev rings $R_{1}$ and $R_{2}$.
The Hamiltonian of the extended Kitaev double rings can be written as $H_{D}=H_{R_1}+H_{R_2}+H_{t}$. $H_{R_1}$ and $H_{R_2}$ are the Hamiltonian of the Kitaev rings $R_1$ and $R_2$, which are described by Eq.~({\ref{eq2}}). The coupling Hamiltonian $H_{t}$ between the two Kitaev rings is given by
\begin{align}
H_{t}=-t_{1}c_{n_1}^{\dag}c_{n^{\prime}_{N_{2}}}-t_{1}c_{n_{N_{1}}}^{\dag}c_{n^{\prime}_{1}}+h.c.,
\label{eq4}
\end{align}
where $c_{n_\xi}^{\dag}$ and $c_{n^{\prime}_{\xi^{\prime}}}$ are creation operator for ring $R_{1}$ and annihilation operator for ring $R_{2}$, respectively. $\xi=1,N_{1}$ ($\xi^{\prime}=1,N_{2}$) are the two coupling sites for the Kitaev ring $R_1$ ($R_{2}$). $t_{1}$ is the coupling strength.

To implement topological braiding, we tune the time-dependent
intensities of initial hopping phase $\phi_{1}^{1,2}$ and hopping phase difference $\delta\phi^{1,2}$, as displayed in Fig.~\ref{fig3}(a).
In Figs.~\ref{fig3}(b)-~\ref{fig3}(f), we calculate the energy levels of the Kitaev double rings with different time $\tau$. The braiding protocol takes four steps in $T$ time to spatially swap two neighboring MBSs $\gamma_{3}$ and $\gamma_{4}$:
(i) At $\tau=0$, we set $\phi_{1}^{1}=0.5\pi$, $\phi_{1}^{2}=0.7\pi$, $\delta\phi^{1}=2\pi$ and $\delta\phi^{2}=0.5\pi$. Figure \ref{fig3}(b) shows that four zero-energy in-gap MBSs (red dots) are present in the extended Kitaev double rings. $\gamma_1$, $\gamma_2$, $\gamma_3$ and $\gamma_4$ arise at four positions of the upper ring $R_1$ [see the inset of Fig.~\ref{fig3}(b)]. These four MBSs can be considered as our initial state in the topological braiding processing.
(ii) At $\tau=T/3$, the initial hopping phase and hopping phase difference change to $\phi_{1}^{1}=\pi$, $\phi_{1}^{2}=0.5\pi$, $\delta\phi^{1}=1.5\pi$ and $\delta\phi^{2}=0.7\pi$. The four zero-energy MBSs remain [see Fig.~\ref{fig2}(c)], but the position of $\gamma_4$ move to ring $R_2$. The remaining three MBSs $\gamma_1$, $\gamma_2$ and $\gamma_3$ are located in the lower left, upper, and lower right sites of the ring $R_1$, respectively [see the inset of Fig.~\ref{fig2}(c)].
(iii) For $\tau \in [T/3, 2T/3]$, we only increase $\phi_{1}^{1}$ of ring $R_1$ from $\pi$ to $2\pi$, and keep the rest of the parameters unchanged. It can be seen that four MBSs remain at $\tau=2T/3$ as shown in Fig.~\ref{fig3}(d).
The inset of Fig.~\ref{fig3}(d) shows that the position of $\gamma_4$ remains unchanged in ring $R_2$, while the positions of $\gamma_1$, $\gamma_2$ and $\gamma_3$ are turned counterclockwise in ring $R_1$ to the upper, lower right, and lower left sites, respectively.
(iv) For $\tau \in [2T/3, T]$, we leave the initial hopping phase of ring $R_1$ unchanged at $\phi_{1}^{1}=2\pi$  and change the remaining three parameters to $\phi_{1}^{2}=0.7\pi$, $\delta\phi^{1}=2\pi$ and $\delta\phi^{2}=0.5\pi$.
One can see from Fig.~\ref{fig3}(e) that the four MBSs still emerge, as shown by red dots.
The inset of Fig.~\ref{fig3}(e) shows that four MBSs $\gamma_1$, $\gamma_2$, $\gamma_3$ and $\gamma_4$ are bounded at the upper left, upper right, lower left and lower right sites of the ring $R_1$, respectively.  The spatial positions of $\gamma_3$ and $\gamma_4$ are mutually swapped, and the spatial positions of $\gamma_1$ and $\gamma_2$ remain unchanged [see the inset of Figs.~\ref{fig3}(b) and \ref{fig3}(e)].
In Video 2 in the Supplementary Material~\cite{supp},
we show the animation of the braiding process when the time $\tau$
increases from $0$ to $T$ as shown in Fig.~\ref{fig3}(a).

In order to observe whether the four MBSs are excited throughout the entire braiding process, we plot the energy levels of the extended Kitaev double rings as a function of the time $\tau$ [see Fig.~\ref{fig3}(f)].
One can see that the zero-energy MBSs (red lines) remain stable throughout the variation of $\tau$.
The isolated zero-energy MBSs can completely prevent mixing with other states (black lines) by an energy gap. Thus, our results validate the stability of the braiding process in the extended Kitaev double rings.
Furthermore, any two adjacent MBSs of the four MBSs can be swapped
in the extended Kitaev double rings.
Thus, a variety of braiding schemes can be designed in a multiple-ring system.

Finally, we present the possibility of an experimental realization of the proposed setup. The phases in Eq.~({\ref{eq1}}) are determined by two initial phases $\phi_1$, $\theta_1$ and two phase differences $\delta\phi$, $\delta\theta$. The value of the initial superconducting phase $\theta_1$ does not affect the appearance and  positions of non-endpoint MBSs.
The gauge transformation $c_n \rightarrow c_{n}e^{i\theta_{n}/2}$ eliminates the phase from the superconducting terms, in exchange for adding the complex amplitude $e^{i(\theta_{n}-\theta_{n+1})/2}$ to the hopping term from the site $n$ to site $n+1$ \cite{Lesser2021}.
The superconducting phase gradient and the initial hopping phase can be converted into each other. The superconducting phase difference $\delta\theta$ and initial hopping phase $\phi_1$ can be tuned by persistent spin current and magnetic flux \cite{Pientka2013, Zha2015, Nava2017, Miao2022}.
Experimentally, a constant force can be introduced around the smaller perimeter of the hopping torus to adiabatically realize the gradient of  $\phi_n$, $F\propto \partial_{t}\phi_{n}$~\cite{Grusdt2014, Yang2020}.
It gives rise to a quantized Hall current perpendicular to the introduced force. Hence, phase gradient can be introduced experimentally to realize phase settings. A lattice-dependent gradient current induces the emergence and movement of non-endpoints MBSs.

\emph{Conclusions}.---In this work, we investigate the energy levels and transport properties of an extended Kitaev chain with a site-dependent linearly increasing hopping phase and $p$-wave superconducting phase.
It is demonstrated that the controlled number of non-endpoint MBSs can be induced in the chain with the hopping phase difference.
Furthermore, we propose a protocol to implement the pairwise exchange of MBSs in an extended Kitaev ring
and the swap of any two MBSs of the multiple MBSs in the extended Kitaev double rings.
Importantly, MBSs remain stable throughout the braiding process,
as evidenced by the fact that the MBSs can be completely protected
from mixing with other states by an energy gap.

This work was financially supported by the National Natural Science Foundation of China (Grant No. 12074097, No. 11921005, and No. 12374034),
the Natural Science Foundation of Hebei Province (Grant No. A2020205013), the Innovation Program for Quantum Science and Technology (Grant No. 2021ZD0302403), and the Strategic Priority Research Program of Chinese Academy of Sciences (Grant No. XDB28000000).

X.Z. and C.-M.M. contributed equally to this work.

\end{document}